\documentclass[12pt]{iopart}
\usepackage{graphicx}

\newcommand{\prhfo}{$\rm Pr_2Hf_2O_7$}
\newcommand{\przro}{$\rm Pr_2Zr_2O_7$}

\begin{document}

\title{Single crystal growth, structure and magnetic properties of~Pr$_{2}$Hf$_{2}$O$_{7}$~pyrochlore}

\author{Monica~Ciomaga~Hatnean$^1$, Romain~Sibille$^2$, Martin~R.~Lees$^1$, Michel~Kenzelmann$^2$, Voraksmy~Ban$^3$, Vladimir~Pomjakushin$^4$, and Geetha~Balakrishnan$^1$}
\address{$^1$Department of Physics, University of Warwick, Coventry, CV4 7AL, UK}
\address{$^2$Laboratory for Scientific Developments and Novel Materials, Paul Scherrer Institut, 5232 Villigen PSI, Switzerland}
\address{$^3$Laboratory for Synchrotron Radiation - Condensed Matter, Paul Scherrer Institut, 5232 Villigen PSI, Switzerland}
\address{$^4$Laboratory for Neutron Scattering and Imaging, Paul Scherrer Institut, 5232 Villigen PSI, Switzerland}

\ead{M.Ciomaga-Hatnean@warwick.ac.uk}
\ead{romain.sibille@psi.ch}

\vspace{10pt}

\begin{abstract}
Large single crystals of the pyrochlore \prhfo~have been successfully grown by the floating zone technique using an optical furnace equipped with high power Xenon arc lamps. Structural investigations have been carried out by both synchrotron X-ray and neutron powder diffraction to establish the crystallographic structure of the materials produced. The magnetic properties of the single crystals have been determined for magnetic fields applied along different crystallographic axes. The results reveal that \prhfo~is an interesting material for further investigations as a frustrated magnet. The high quality of the crystals produced make them ideal for detailed investigations, especially those using neutron scattering techniques. 
\end{abstract}

Accepted for publication in {\JPCM}.
%
%
%
%
%

\section{\label{sec:level1}Introduction}

Pyrochlore oxides have been a topic of great interest due to the intriguing magnetic properties that originate from their frustrated magnetic lattice~\cite{Blote1969,Bramwell2001,Greedan2001,Gardner2010,Malkin2010,Petrenko2011,Gingras2014}. Compounds of the general formula $A_{2}B_{2}$O$_{7}$ (where $A$ and $B$ are metals), the pyrochlores, have a face-centered cubic structure with space group $Fd\bar{3}m$ (No. 227). The majority of the pyrochlore oxides belong to the (3+, 4+) type $A_{2}^{3+}B_{2}^{4+}$O$_{7}$, where the $A$ sites are occupied by trivalent cations located in the centre of scalenohedra (distorted cubes) of oxygen anions, and the $B$ sites are occupied by tetravalent transition metal ions situated in the centre of oxygen octahedra~\cite{Subramanian1983}. $A$ and $B$ cations form, independently, two so-called pyrochlore lattices of corner-sharing tetrahedra~\cite{Greedan2001,Gardner2010}. Depending on the nature and strength of the magnetic moment and interactions, the pyrochlore oxides can display a wide variety of magnetic ground states~\cite{Gardner2010} ranging from spin ice~\cite{Harris1997,Bramwell2001,Bramwell2001b} where the spin correlations lead to a Coulomb phase~\cite{Henley:2010vo,Fennell2009} with emergent magnetostatics~\cite{Castelnovo:2008hb,Castelnovo:2012kk}, spin frozen states,\cite{Greedan1986,Gaulin1992,Zhou2008} to long-range ordered states (see \cite{Gardner2010} and references therein).

Additionally, one of the most intriguing areas of current research is into materials which exhibit Quantum Spin Liquid (QSL) ground states, as has been pointed out in recent studies of pyrochlores based on Yb$^{3+}$~\cite{Ross:2011tv,Chang:2012el,Applegate2012}, although this case is under debate \cite{Robert2015,Jaubert2015}, Pr$^{3+}$~\cite{Zhou:2008cz,Matsuhira2009,Onoda2010,Lee2012,Kimura2013} or Ce$^{3+}$~\cite{Sibille2015}. The candidate materials for the realization of QSL states are based on the rare-earth ions that develop relatively small magnetic moments. The reason for this is that, for small magnetic moments, the transverse terms in the effective spin $-1/2$ Hamiltonian on the pyrochlore lattice~\cite{Curnoe2008,Ross2011}, which are responsible for the stabilization of quantum phases~\cite{Gingras2014,Matsuhira2009,Hermele2004,Benton2012,Savary2012}, are not overwhelmed by the dipolar interaction which leads to the classical spin ice when the dipolar interaction dominates.

The synthesis of large, high quality single crystals of pyrochlore oxides, and in particular the rare-earth titanates~\cite{Balakrishnan1998,Gardner1998,Prabhakaran2011}, some rare-earth molybdates,~\cite{Taguchi2002,Kezsmarki2004} and zirconates~\cite{Matsuhira2009,CiomagaHatnean2014,Koohpayeh2014,CiomagaHatnean2015} has been accomplished by the floating-zone technique. This success has allowed real and rapid progress to be made in the investigation of frustrated magnets and to unearth some very interesting magnetic properties. All the members of the titanate pyrochlore family have been thoroughly investigated over the past years, whilst the molybdate~\cite{Taguchi2002,Kezsmarki2004,Hanasaki2006,Kezsmarki2006} and the zirconate~\cite{Matsuhira2009,Kimura2013,Kimura2013_1,CiomagaHatnean2014,CiomagaHatnean2015_1} series have only recently come to the attention of the research community.
Recent studies report on the availability of large crystals of the frustrated pyrochlore magnet Nd$_{2}$Hf$_{2}$O$_{7}$~\cite{Chun2015}. The floating zone technique is ideal to produce crystals of other members of the rare-earth hafnate pyrochlores. This is particularly appealing since the structural and magnetic characteristics of the hafnate family have not yet been investigated in great detail.

Recent studies of the intriguing magnetic properties of the praseodymium based pyrochlores, \przro~\cite{Kimura2013} and Pr$_{2}$Ir$_{2}$O$_{7}$~\cite{Tokiwa2014} have motivated us to embark upon the study of the analogous compound in the hafnate pyrochlore series, \prhfo. We have succeeded in preparing, for the first time, single crystals of the praseodymium hafnate pyrochlore by the floating-zone technique. The growth of large high-quality single crystals of this oxide represents an important step in the field, and opens up a route to further investigations of this novel class of pyrochlores, with the potential to lead to an in-depth understanding of the effects of frustration in praseodymium containing pyrochlores~\cite{Onoda2010}. In this paper, we report the synthesis, the structural characterization and preliminary studies of the magnetic properties of single crystals of the praseodymium hafnate pyrochlore, \prhfo.

\section{Experimental section}

Polycrystalline samples of \prhfo~were prepared by conventional solid state reaction. Stoichiometric quantities of the starting materials, Pr$_{6}$O$_{11}$ (Chempur, 99.999\%) and HfO$_{2}$ (Chempur, 99.95\%), were mixed, ground and heated to 1300~$^{\circ}$C for 10 hours and 1550~$^{\circ}$C for 10 hours with intermediate grinding. The synthesized powder was thoroughly reground and then isostatically pressed into cylindrical rods, 6-8~mm in diameter and about 60-70~mm long. The resulting rods were sintered for several days in air at 1450~$^{\circ}$C in preparation for crystal growth experiments.

Single crystals of \prhfo~were grown by the floating-zone technique using a four-mirror xenon arc lamp optical image furnace (CSI FZ-T-12000-X\_VI-VP, Crystal Systems, Inc., Japan). The growths were performed in high purity argon at a pressure of $\sim$2~bars, using a growth rate of 18~mm/h. The feed and the seed rods were counter-rotated at around $\sim$20-30~rpm. Initially, a crystal boule of \przro\ was used as seed and once good quality crystals of \prhfo\ were obtained, the subsequent growths were carried out using crystal boules of \prhfo\ as seeds.

Powder X-ray diffraction experiments were carried out at the Swiss Light Source (SLS) using the MS beamline (powder station) \cite{JSY2:JSYIE5093}. A diffraction pattern of the \prhfo\ polycrystalline material (starting material for the growth) was measured in a quartz capillary ($\phi = 0.1$~mm) with the Debye-Scherrer geometry and a multistrip MYTHEN II detector. The incident beam had an energy of $\sim22$~keV ($\lambda=0.564941$~\AA) and the diffracted beams were measured up to $60^{\circ}$ $2\theta$, with a step size of $\sim0.0036^{\circ}$ 2$\theta$. Powder neutron diffraction experiments were carried out on the starting polycrystalline material at the Swiss spallation neutron source SINQ using the HRPT diffractometer ($\lambda=1.155$~\AA). Diffraction was measured between 5 and $162^{\circ}$ 2$\theta$, with a step size of $0.05^{\circ}$ in 2$\theta$, in a standard "Orange" helium cryostat. A joint Rietveld~\cite{Rietveld1969} refinement with equal weighting factors for the synchrotron and neutron data was performed using the \textsc{fullprof} software suite~\cite{RodriguezCarvajal1993}. The instrumental resolution functions were experimentally determined from the measurements of small linewidth standards. A total of 23 parameters were refined: two sets of parameters independently refined for the two diffraction patterns (scale factors, zero-shifts, lattice parameter, sample contributions to the peak shapes), and twelve parameters defining the structural model ($x$ coordinate of the 48\textit{f} oxygen atom, anisotropic displacement parameters of all atoms, occupancy factors of three atomic positions).

In addition to the powder synchrotron X-ray diffraction pattern measured on the starting polycrystalline material, a measurement was also performed on a sample obtained by grinding a tiny crystal fragment from the middle of a single crystal. The drawbacks of using a powdered crystal fragment were the small amount of the resulting powder and the difficulty in grinding the crystal sample into very fine particles. The aforementioned factors hindered the realization of a full Rietveld analysis, because of insufficient powder averaging and poorly modelled lineshapes. Instead, that data was analyzed by a Le Bail decomposition~\cite{LeBail1988} that was good enough to provide a precise estimate of the lattice parameter. The agreement factors for the refinement of the powder diffraction data given in the manuscript are defined in Ref.~\cite{McCusker1999}.

A Laue X-ray imaging system with a Photonic-Science Laue camera was used to investigate the quality of the crystal boules and to orient single-crystal samples for selected experiments. A rectangular-prism shaped sample, with dimensions of $2.73\times1.81\times1.96$~mm$^{3}$, was cut from the \prhfo\ boule for magnetization measurements. The sample was cut in order that the [110] (rhombic) and [001] (tetragonal) directions were perpendicular to the faces of the rectangular prism. The demagnetizing factors were calculated using expressions derived by Aharoni~\cite{Aharoni1998}. 

Magnetization measurements were carried out using a Quantum Design Magnetic Property Measurement System MPMS-5S superconducting quantum interference device (SQUID) magnetometer along with an i-Quantum $^{3}$He insert. The magnetic susceptibility, which is equal to the magnetization $M$ divided by the magnetic field $H$ in the linear field regime, was evaluated as a function of temperature in a constant applied magnetic field of 1~kOe from 0.5 to 300~K. Magnetization measurements were also performed as a function of magnetic field up to 70 kOe directed along specific crystallographic axes at various temperatures.

\section{Results and Discussion}

\subsection{Crystal chemistry}

\begin{figure}[ht]
\begin{center}
\includegraphics[width=4in]{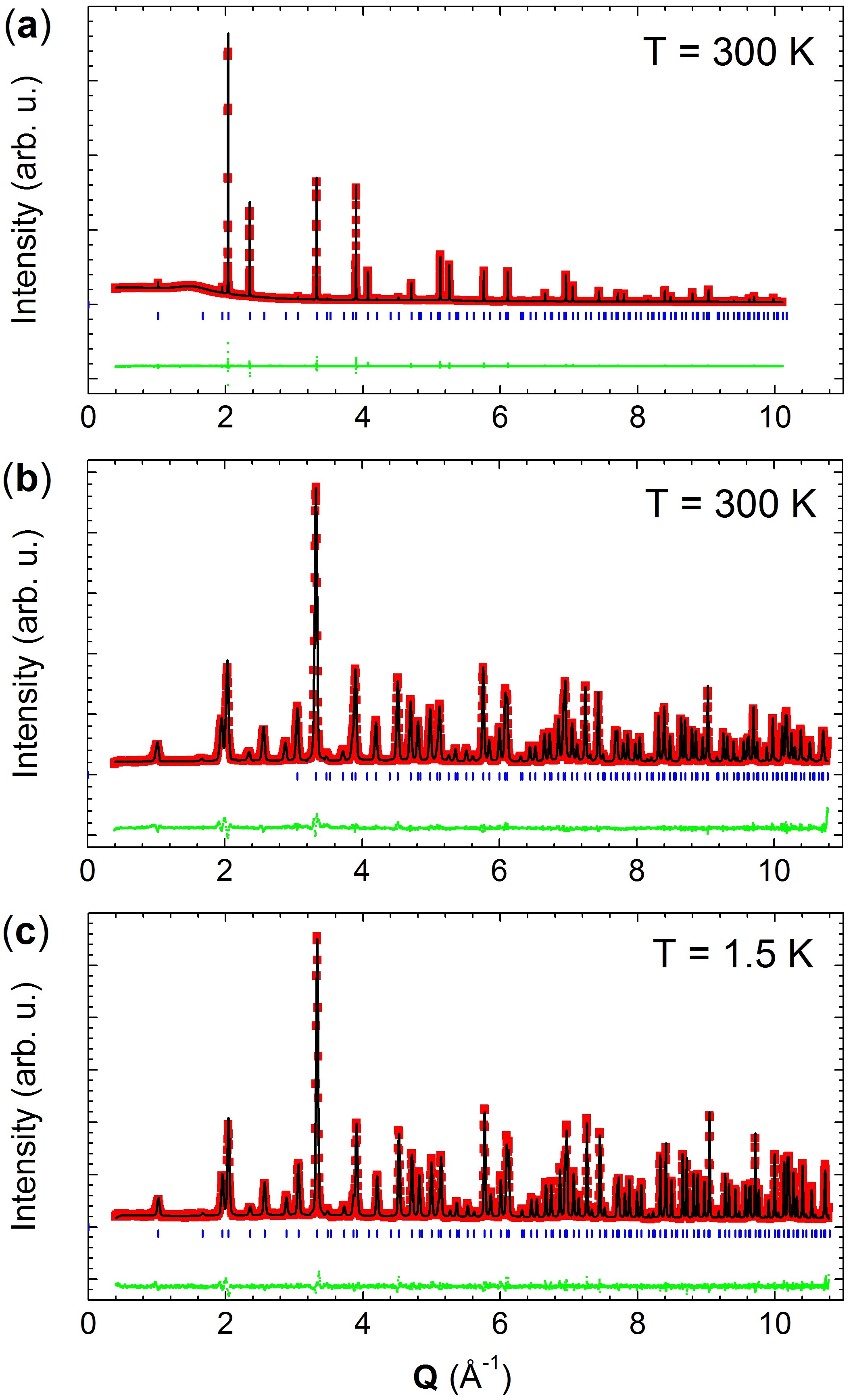}
\caption{\label{fig1diff} Rietveld refinement of powder (a) synchrotron X-ray and (b and c) neutron diffraction data collected at 300~K (a and b) and at 1.5~K (c). Black, red, and green represent the experimental data, the fit, and difference between the data and the fit respectively, while the blue ticks indicate the Bragg positions. The incident wavelengths are $\lambda=0.621418$~\AA~(a) and $\lambda=1.155$~\AA~(b and c). Patterns (a) and (b) are jointly refined against a common structural model given in Table~\ref{crystal_data}. The low-temperature structure corresponding to pattern (c) is presented in Table~\ref{crystal_data_LT}. Conventional Rietveld factors for pattern (a) (\%): $R_\mathrm{{P}}=1.70$; $R_\mathrm{{WP}}=2.04$; $R_\mathrm{{Bragg}}=2.66$; $R_\mathrm{F}=7.45$. Conventional Rietveld factors for pattern (b) (\%): $R_\mathrm{{P}}=3.88$; $R_\mathrm{{WP}}=5.19$; $R_\mathrm{{Bragg}}=4.24$; $R_\mathrm{F}=2.58$. Conventional Rietveld factors for pattern (c) (\%): $R_\mathrm{{P}}=3.88$; $R_\mathrm{{WP}}=5.19$; $R_\mathrm{{Bragg}}=3.91$; $R_\mathrm{F}=2.55$.}
\end{center}
\end{figure}

Firstly, we have investigated in detail the crystal structure of \prhfo\ in order to confirm the relevance of this material as a model pyrochlore magnet. We have used the polycrystalline sample prepared as starting material for the crystal growth for these experiments. Diffraction patterns were measured using synchrotron X-ray (Fig.~\ref{fig1diff}a) and neutron (Fig.~\ref{fig1diff}b) radiation and refined together against the pyrochlore structure (space group $Fd\bar{3}m$, origin choice 2). The Rietveld procedure converges rapidly and the conventional agreement factors for Rietveld refinements~\cite{McCusker1999} are $R_\mathrm{{WP}}=2.04$ and $R_\mathrm{{Bragg}}=2.66$, and $R_\mathrm{{WP}}=5.19$ and $R_\mathrm{{Bragg}}=4.24$, respectively for the synchrotron X-ray and neutron patterns at 300~K. 

\begin{figure}[ht]
\begin{center}
\includegraphics[width=\linewidth]{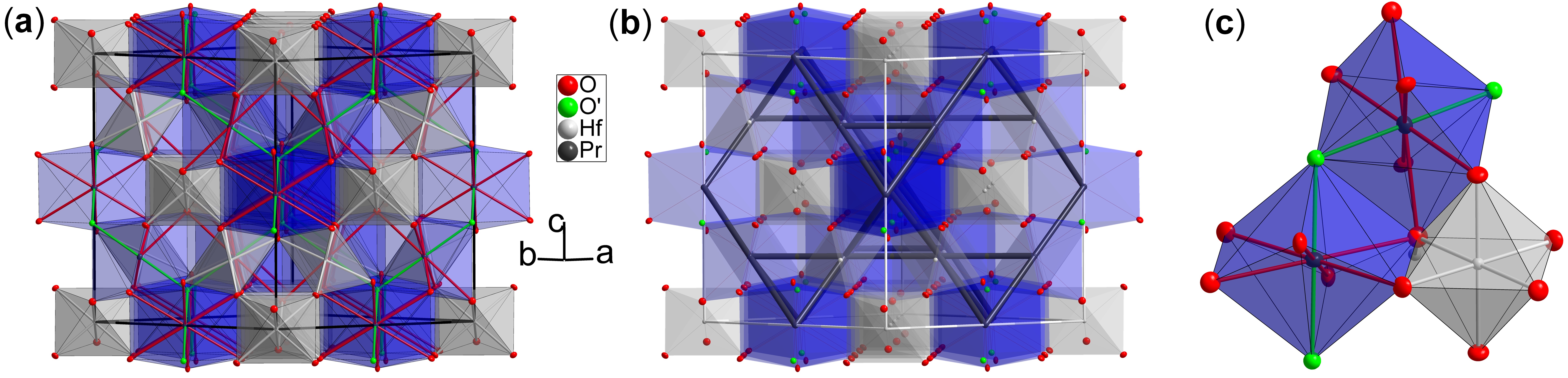}
\caption{\label{figstruct} Crystal structure of \prhfo\ obtained from the joint Rietveld refinement of synchrotron X-ray and neutron data measured at room temperature, $Fd\bar{3}m$, origin choice 2. In panel (a) the cell edges are drawn in black, and we emphasize the Pr-O (red), Pr-O$^\prime$ (green) and Hf-O (grey) bonds. Blue and grey polyhedra show the oxygen scalenohedra and octahedra around Pr$^{3+}$ and Hf$^{4+}$ cations, respectively. In panel (b) the cell edges are drawn in white and we emphasize the magnetic pyrochlore lattice of Pr$^{3+}$ cations (in black). Panel (c) shows the local coordination around the two metals and the connection between the polyhedra.}
\end{center}
\end{figure}

The joint refinement provides a complete description of the room-temperature crystal structure: precise lattice parameter, bond distances/angles and anisotropic displacement parameters (ADPs), as well as a certain degree of sensitivity to the chemical composition thanks to the strong contrast in the neutron scattering lengths ($b_{\mathrm{Pr}}=4.58(5)$~fm, $b_{\mathrm{Hf}}=7.77(14)$~fm, $b_{\mathrm{O}}=5.805(4)$~fm). The results are summarized in Table~\ref{crystal_data} and are in good agreement with previously published data~\cite{Karthik2012,Blanchard2013}. The resulting crystal structure at 300~K is shown on Fig.~\ref{figstruct}.
The lattice parameter obtained from the powder synchrotron X-ray data is $10.68411(2)$~\AA. The value of the atomic coordinate $\it{x}$ for the oxygen atom O(48$\it{f}$) is 0.33247(7), in the range of the typical values for $A_{2}B_{2}$O$_{7}$ compounds~\cite{Gardner2010}. The Pr-O(48$f$) bond length is $2.6016(6)$~\AA, close to the sum of the ionic radii ($\sim2.66$ \AA), while Pr-O$^\prime$(8$b$) bond (pointing along the local $\langle111\rangle$ direction) has a length of $2.312697(6)$ \AA, which is markedly shorter than the $2.66$~\AA~usually observed in rare-earth pyrochlores. Attempts to refine antisite cation disorder and oxygen Frenkel disorder did not provide evidence for any deviation from a perfectly ordered pyrochlore structure.

The joint refinement can also be used to retrieve information concerning the composition of the sample. Three of the four occupancy factors were refined in order to avoid total correlation with the scale factor. The refined chemical occupancies remain very close to unity (see Table~\ref{crystal_data}). Given this result it is reasonable to assume a stoichiometric formula \prhfo\ for our polycrystalline material.

\begin{table}[tb]
\centering
\caption{Structural parameters for a polycrystalline sample of \prhfo\ determined from a joint Rietveld refinement of synchrotron X-ray and neutron diffraction data measured at 300~K (space group $Fd\bar{3}m$, origin choice 2).}
\label{crystal_data}
\begin{tabular}{l|llllll}
\hline
$\textbf{\emph{T} = 300}$~\textbf{K}  & \multicolumn{1}{c}{\textit{x}} & \multicolumn{1}{c}{\textit{y}} & \multicolumn{1}{c}{\textit{z}} & \multicolumn{1}{c|}{} & \multicolumn{1}{c}{Occupancy}    &     \\ \hline
Pr (16\textit{d})  & \multicolumn{1}{c}{0.5}        & \multicolumn{1}{c}{0.5}        & \multicolumn{1}{c}{0.5}        & \multicolumn{1}{c|}{} & \multicolumn{1}{c}{0.994(12)} &     \\
Hf (16\textit{c})  & \multicolumn{1}{c}{0}          & \multicolumn{1}{c}{0}          & \multicolumn{1}{c}{0}          & \multicolumn{1}{c|}{} & \multicolumn{1}{c}{1} &     \\
O (48\textit{f})  & \multicolumn{1}{c}{0.375}      & \multicolumn{1}{c}{0.375}      & \multicolumn{1}{c}{0.375}      & \multicolumn{1}{c|}{} & \multicolumn{1}{c}{1.024(15)} &     \\
O$^\prime$ (8\textit{b})   & \multicolumn{1}{c}{0.33247(7)} & \multicolumn{1}{c}{0.125}      & \multicolumn{1}{c}{0.125}      & \multicolumn{1}{c|}{} & \multicolumn{1}{c}{1.012(27)}        &     \\ \hline
 ADPs in~\AA$^2$:    & $\textit{U}_{11}$      & $\textit{U}_{22}$     & $\textit{U}_{33}$        & $\textit{U}_{12}$    & $\textit{U}_{13}$   & $\textit{U}_{23}$ \\ \hline
Pr (16\textit{d})  & 0.00702           & 0.00702            & 0.00702               & -0.00084          & -0.00084          & -0.00084   \\
Hf (16\textit{c})  & 0.00355            & 0.00355            & 0.00355             & 0.00015                     & 0.00015             & 0.00015   \\
O1 (48\textit{f})  & 0.00966          & 0.00712            & 0.00712               & 0                     & 0                           & 0.00269   \\
O2 (8\textit{b})   & 0.00665                & 0.00665         & 0.00665            & 0                     & 0                           & 0  
\end{tabular}
\end{table}

Finally, we have also collected a powder neutron diffraction pattern of \prhfo\ at 1.5~K, in order to determine values for the bond distances and angles that are relevant for the low-temperature superexchange pathways, and to check for structural distortions that may affect the magnetism of the non-Kramers $\rm Pr^{3+}$ ions. The crystal structure maintains its cubic symmetry at 1.5~K, where the lattice ($a=10.66564(5)$~\AA) contracts by about 1.5\% compared to its value at 300~K ($a=10.68189(5)$~\AA\ obtained from the refinement of the neutron diffraction pattern). The results of the refinement at 1.5~K which is presented in Fig.~\ref{fig1diff}c, are summarized in Table~\ref{crystal_data_LT}. At 1.5~K the Pr-O(48$f$) bond length is $2.5937(6)$~\AA\ and the Pr-O$^\prime$ (8$b$) bond has a length of $2.309179(6)$~\AA.

\begin{table}[tb]
\centering
\caption{Structural parameters for a polycrystalline sample of \prhfo\ obtained from a Rietveld refinement of powder neutron diffraction data measured at 1.5~K  (space group $Fd\bar{3}m$, origin choice 2). Results are given assuming all the occupancy factors are equal to unity.}
\label{crystal_data_LT}
\begin{tabular}{l|llllll}
\hline
$\textbf{\emph{T} = 1.5}$~\textbf{K} & \multicolumn{1}{c}{\textit{x}} & \multicolumn{1}{c}{\textit{y}} & \multicolumn{1}{c}{\textit{z}} & \multicolumn{1}{c}{} & \multicolumn{1}{c}{} &     \\ \hline
Pr (16\textit{d})   & \multicolumn{1}{c}{0.5}        & \multicolumn{1}{c}{0.5}        & \multicolumn{1}{c}{0.5}        & \multicolumn{1}{c}{} & \multicolumn{1}{c}{} &     \\
Hf (16\textit{c})  & \multicolumn{1}{c}{0}          & \multicolumn{1}{c}{0}          & \multicolumn{1}{c}{0}          & \multicolumn{1}{c}{} & \multicolumn{1}{c}{} &     \\
O (48\textit{f})  & \multicolumn{1}{c}{0.375}      & \multicolumn{1}{c}{0.375}      & \multicolumn{1}{c}{0.375}      & \multicolumn{1}{c}{} & \multicolumn{1}{c}{} &     \\
O$^\prime$ (8\textit{b})   & \multicolumn{1}{c}{0.33301(7)} & \multicolumn{1}{c}{0.125}      & \multicolumn{1}{c}{0.125}      & \multicolumn{1}{c}{} & \multicolumn{1}{c}{} &     \\ \hline
 ADPs in~\AA$^2$:    & $\textit{U}_{11}$      & $\textit{U}_{22}$     & $\textit{U}_{33}$        & $\textit{U}_{12}$    & $\textit{U}_{13}$   & $\textit{U}_{23}$ \\ \hline
Pr (16\textit{d})  & 0.00334           & 0.00334            & 0.00334               & 0.00062          & 0.00062          & 0.00062   \\
Hf (16\textit{c})  & 0.00116            & 0.00116            & 0.00116             & -0.00015                     & -0.00015             & -0.00015   \\
O1 (48\textit{f})  & 0.00480          & 0.00460            & 0.00460               & 0                     & 0                           & 0.00008   \\
O2 (8\textit{b})   & 0.00419                & 0.00419         & 0.00419            & 0                     & 0                           & 0  
\end{tabular}
\end{table}

\subsection{Crystal growth}

\begin{figure}[ht]
\begin{center}
\includegraphics[width=6in]{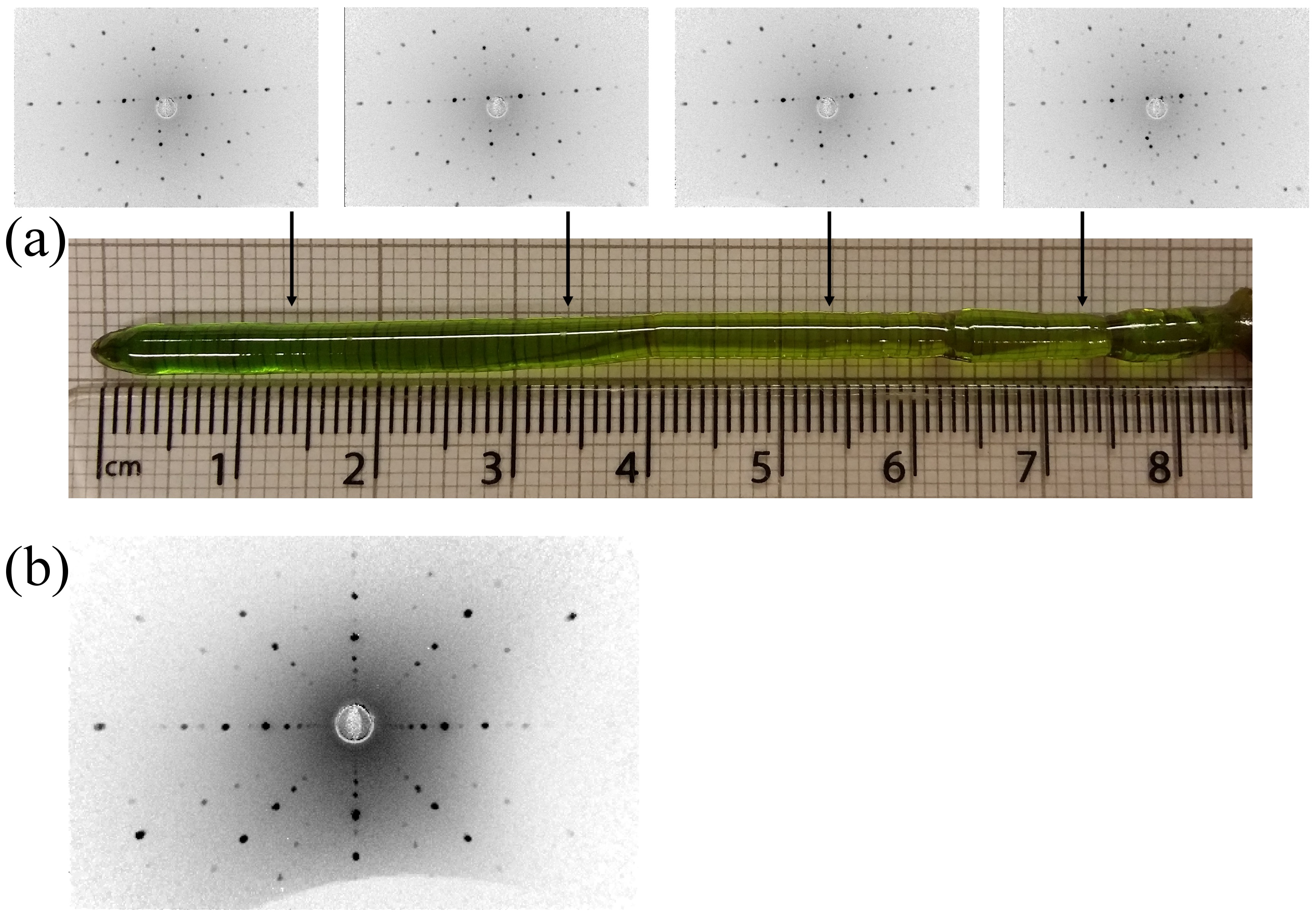}
\caption{\label{prhfo_crystal} (a) Crystal of \prhfo\ grown in high purity argon atmosphere, at a pressure of $\sim$2~bars and a translation rate of 18~mm/h. Also shown above the image of the crystal are the Laue patterns of one of the facets, taken along the crystal length at $\sim$2~cm intervals, between the end (left) and the beginning (right) of the boule. The corresponding Laue patterns taken on the facet at 180~degrees are mirror images of these patterns. (b) Laue back reflection X-ray photograph of an aligned sample (showing the [001] orientation) used for the magnetic properties measurements discussed in the present work.}
\end{center}
\end{figure}

Crystals of \prhfo\ were successfully grown by the floating zone method, using similar growth conditions to those used for preparing \przro\ crystal boules~\cite{CiomagaHatnean2014,CiomagaHatnean2016}. One of the difficulties associated with the growth of praseodymium related compounds is the evaporation of Pr$_{2}$O$_{3}$ during the crystal growth process, which can cause a decrease in the Pr content in the single crystals~\cite{Matsuhira2009,CiomagaHatnean2014,CiomagaHatnean2016}. This phenomena can be avoided by employing a high growth rate and by performing the growth in a pressurized gas atmosphere (inside a quartz tube) to suppress the evaporation~\cite{CiomagaHatnean2014}. The crystal growth of \prhfo\ was performed in high purity argon gas in order to facilitate the reduction of the Pr$^{4+}$ ions to Pr$^{3+}$ (see Refs.~\cite{CiomagaHatnean2014} and ~\cite{Koohpayeh2014}, and references therein). 
\prhfo\ crystals obtained were typically 5-7~mm in diameter and 60-85~mm long. The crystals developed well defined facets, within the first few millimeters of the growth and the boules obtained were free of any cracks. No deposition was observed on the quartz tube surrounding the sample during the growth process, suggesting that no evaporation occurred during any of the growths.  All the praseodymium hafnate boules were transparent to light, with a bright green colour. A photograph of an as-grown crystal of \prhfo\ is shown in Fig.~\ref{prhfo_crystal}a. The crystal quality of the boules was investigated by Laue X-ray diffraction, and Laue photographs were taken along the length of the boule, on the faceted sides (see Fig.~\ref{prhfo_crystal}a). The Laue patterns were identical along the whole length of the faceted faces and, in most cases, the [110] direction is almost orthogonal to one of the facets. A Laue photograph taken on an aligned sample of \prhfo\ used for magnetic properties measurements is shown in Fig.~\ref{prhfo_crystal}b.

\begin{figure}[ht]
\begin{center}
\includegraphics[width=6in]{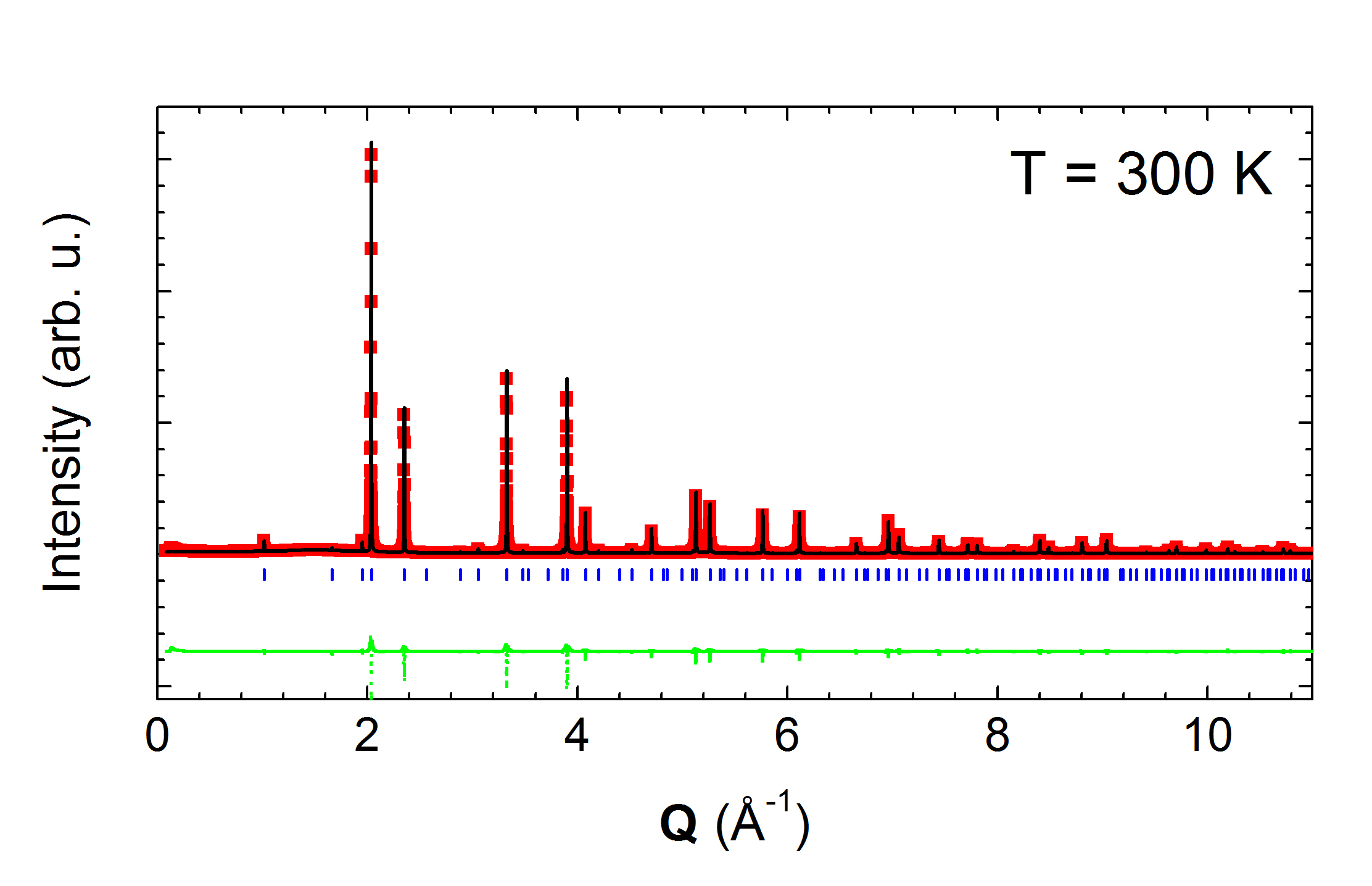}
\caption{\label{prhfo_XRD} Powder synchrotron X-ray diffraction pattern collected on a ground crystal of \prhfo\ at 300 K. The experimental profile (red) and a Le Bail decomposition (black) are shown, with the difference given in green. The Bragg positions are indicated by the blue ticks.}
\end{center}
\end{figure}

The powder synchrotron X-ray diffraction pattern of a ground fragment taken from the middle of the specimen shown in Fig.~\ref{prhfo_crystal}a was refined against the pyrochlore lattice (Fig.~\ref{prhfo_XRD}). The pattern matches very well with the cubic pyrochlore phase and no impurity peaks were present. Furthermore, the superlattice reflections which are the characteristic trademarks of the pyrochlore structure are clearly visible in the X-ray diffraction pattern. The lattice parameter (10.67704(3)~\AA) was found to be slightly smaller than the value of $10.68411(2)$~\AA\ obtained at the same temperature and using the same method for our polycrystalline material (Fig.~\ref{fig1diff}a). A difference in the value of the lattice parameter between the polycrystalline and single crystal samples has also been reported for the \przro\ pyrochlore~\cite{CiomagaHatnean2014}. The smaller lattice parameter observed in the single crystals may be attributed to a very small difference in the stoichiometry of polycrystalline and single crystalline samples. We note, however, that we could not find evidence for different physical behaviour in powder and single crystal samples, which appears consistent with the good agreement between our heat capacity data taken on single crystal samples~\cite{Sibille2016} and other data recently reported for powder samples~\cite{Anand2016}. 

\subsection{Magnetic properties}

\begin{figure}[ht]
\begin{center}
\includegraphics[width=6in]{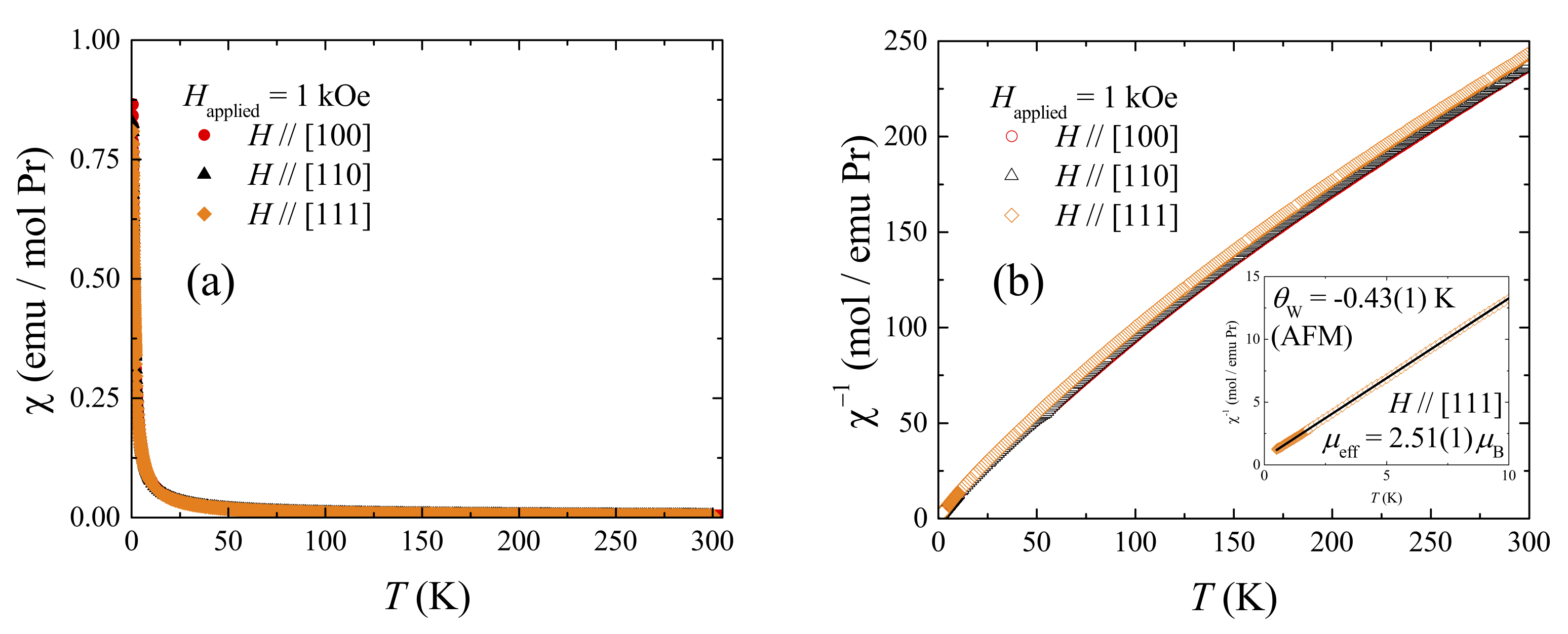}
\caption{\label{prhfo_chivT} (a) Temperature dependence of the \textit{dc} magnetic susceptibility, $\chi$ versus $T$, in the temperature range 0.5 to 300~K for a crystal of \prhfo, with a magnetic field applied along the [100] (red), [110] (black), and [111] (orange) directions. (b) Temperature dependence of the reciprocal of the bulk $dc$ susceptibility, $\chi^{-1}$ versus $T$, for a field applied along the three high symmetry directions. The inset shows $\chi^{-1}$ versus $T$ and the linear fit (using the Curie-Weiss law) to the data in the temperature range 0.5 to 10~K for a magnetic field applied along the [111] direction.}
\end{center}
\end{figure}

Field-cooled (FC) and zero-field-cooled (ZFC) magnetization versus temperature data were collected on a \prhfo\ rectangular-prism-shaped single crystal aligned along the three high symmetry crystallographic directions ([100], [110] and [111]). The data were corrected for demagnetization effects~\cite{Aharoni1998}; the demagnetizing factors were found to be equal to $N = 0.38$, 0.35 and 0.25, respectively, where $H=H_{\mathrm{applied}}-4\pi MN$. Fig.~\ref{prhfo_chivT} shows the temperature dependency of the \textit{dc} magnetic susceptibility, $\chi\left(T\right)$, and the reciprocal \textit{dc} magnetic susceptibility $\chi^{-1}\left(T\right)$. The data measured along the different crystallographic axes in 1~kOe reveal a monotonic and highly-isotropic (within experimental error) increase upon cooling from $T=300$ to 0.5~K, and the absence of any anomaly which may indicate a magnetic transition. $\chi^{-1}\left(T\right)$ data do not obey a Curie-Weiss law in the temperature range 0.5 to 300~K, although fits could be made over a reduced temperature range (0.5 to 10~K) [see Fig.~\ref{prhfo_chivT}(b)(inset)]. It was found that the results of the fits depend on the exact temperature range over which the fit is performed. These results highlight the importance of investigating the crystal electric field (CEF) scheme in these systems. The crystal field splitting of Pr$^{3+}$ in \prhfo\ was determined and the results are described elsewhere~\cite{Sibille2016}. In \prhfo\ the first excited level is about 9.2 meV $\sim107$~K above the ground state doublet, meaning that information concerning the magnetic interactions can be deduced from a Curie-Weiss analysis well below this temperature. Accordingly, a fit of the magnetic susceptibility to a Curie-Weiss law was made in the temperature range from 0.5 to 10~K, yielding a Curie-Weiss temperature of $\theta_\mathrm{{W}}=-0.43(1)$~K for the magnetic field applied along the [111] direction. This indicates the presence of antiferromagnetic interactions, slightly weaker than in \przro\ where $\theta_\mathrm{{W}}=-1.4(1)$~K~\cite{Kimura2013}. The Pr$^{3+}$ effective moment is estimated to be $\mu_\mathrm{{eff}}=2.51(1)\mu_\mathrm{{B}}$, a similar value to the one found in \przro~\cite{Kimura2013,CiomagaHatnean2014}. The calculated values of the Curie-Weiss temperature $\theta_\mathrm{{W}}$, and of the effective moment $\mu_\mathrm{{eff}}$ are in agreement with those determined by detailed investigations of the low temperature magnetic properties of our \prhfo\ crystals~\cite{Sibille2016}. We note the apparent discrepancy between the negative Curie-Weiss temperature and the fact that the physics of Pr-based pyrochlores appears to be related to spin ice~\cite{Kimura2013,Sibille2016,Petit2016}. However, it has recently been established using inelastic neutron spectroscopy~\cite{Petit2016} that, in \przro, the parameters of the Hamiltonian for Pr-based pyrochlores~\cite{Onoda2010} lead to a phase where quadrupolar correlations can overcome the antiferromagnetic exchange and account for the spin ice-like structure factor.

\begin{figure}[ht]
\begin{center}
\includegraphics[width=6in]{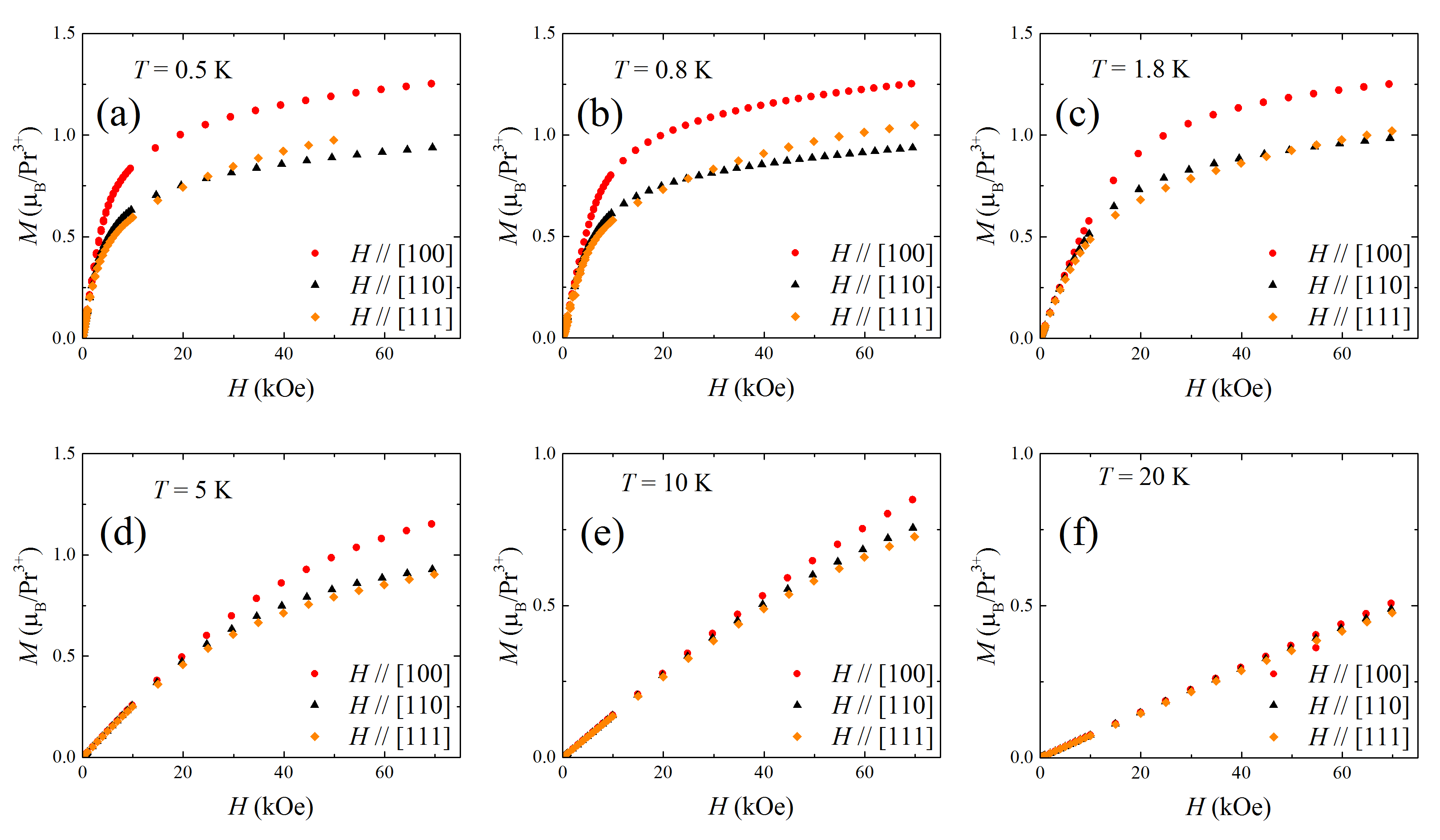}
\caption{\label{prhfo_MvH} Isothermal magnetization ($M$) as a function of applied magnetic field ($H$) along the [100] (red), [110] (black), and [111] (orange) directions at temperatures of (a) 0.5, (b) 0.8, (c) 1.8,  (d) 5, (e) 10 and (f) 20~K for a single-crystal of \prhfo.}
\end{center}
\end{figure}

The magnetization measured along the three directions as a function of applied magnetic field $M(H)$ at various temperatures is shown in Fig.~\ref{prhfo_MvH}. The data collected at 1.8~K or below [see Figs.~\ref{prhfo_MvH}(a) and \ref{prhfo_MvH}(b)] reveal a non linear response of the magnetization as the applied field increases. Furthermore, the field dependence of the magnetization appears reversible, with no hysteresis between the field increasing and field decreasing $M(H)$ curves. The magnetization measured with a magnetic field applied along the [100] direction is the highest in strong magnetic fields, whilst the [110] direction gives the lowest values of the magnetization. These results suggest similarities in terms of local anisotropy of the magnetic moments between the \prhfo\ pyrochlore and the classical spin ice systems such as Dy$_{2}$Ti$_{2}$O$_{7}$ and Ho$_{2}$Ti$_{2}$O$_{7}$~\cite{Fukazawa2002,Petrenko2003}. However, the values of the magnetic moments measured for the three crystallographic directions at the maximum applied field are smaller than the values of the expected saturated moments for a classic spin ice configuration~\cite{Fukazawa2002}. A similar local $\langle111\rangle$ Ising behaviour was also observed in the related praseodymium zirconate pyrochlore, \przro~\cite{Kimura2013,Kimura2013_1,CiomagaHatnean2014}. In \prhfo\ the magnetization response is temperature dependent (see Figs.~\ref{prhfo_MvH}(a)-(f)), with the strongly anisotropic response only observed below about 10~K. Finally, we note the absence of evidence for a magnetization plateau developing in the $M$($H$) curve when the field is applied along the [111] direction, as opposed to observations made at much lower temperature~\cite{Sibille2016}. The discrepancy only arises from the difference in temperature, because ferromagnetic correlations only develop below 0.5~K in this system~\cite{Sibille2016}.

\section{Summary}

We have successfully prepared large, high-quality single crystals of the novel frustrated pyrochlore magnet \prhfo\ by the floating zone technique, using a growth rate of 18~mm/h in a high purity argon atmosphere, at a pressure of $\sim$2~bars. Powder X-ray diffraction studies confirm that the crystal boules are of single-phase pyrochlore $Fd\bar{3}m$ structure. The quality of our \prhfo\ single crystals appears to be very high according to several criteria (colour homogeneity, absence of cracks, transparency, quality and spatial homogeneity of the X-ray Laue diffraction patterns). The temperature dependence of the magnetic susceptibility measured in a low magnetic field shows an isotropic behaviour without any sign of long-range magnetic ordering down to 0.5~K. The field dependence of the isothermal magnetization reveals an anisotropic behaviour at low temperature, indicating a spin ice type of anisotropy. This magnetic response is similar to that seen in the related pyrochlore praseodymium zirconate, \przro. However, in contrast to the \przro\ pyrochlore in which recent results point to the existence of a certain degree of disorder~\cite{Petit2016}, the investigations performed on our \prhfo\ crystals show a structure with no cationic or anionic deficiencies. A recent determination of the crystal field scheme in \prhfo\ using neutron spectroscopy on polycrystalline samples~\cite{Sibille2016} has confirmed the nature of the anisotropy deduced for the bulk measurements presented here. The \prhfo\ crystals produced are ideal for further investigations of the low temperature magnetism by neutron scattering techniques.

\section*{Acknowledgments}
This work was supported by a grant from the EPSRC, UK (Grant No. EP/M028771/1). The authors thank Mr. Tom E. Orton for valuable technical support.  We acknowledge funding from the European Community's Seventh Framework Program (Grant No. 290605, COFUND: PSI-FELLOW), and the Swiss National Science Foundation (Grant Nos. 200021\_138018 and 200020\_162626). Neutron scattering experiments were carried out at the continuous spallation neutron source SINQ at the Paul Scherrer Institut at Villigen PSI in Switzerland and at the Institut Laue Langevin in Grenoble, France. Synchrotron powder X-ray diffraction measurements were carried out at the Materials Science beamline X04SA of the Swiss Light Source (SLS) at the Paul Scherrer Institut at Villigen PSI in Switzerland.

\section*{References}
\bibliographystyle{iopart-num}
\bibliography{Pr2Hf2O7.bib}

\providecommand{\newblock}{}
\begin{thebibliography}{10}
\expandafter\ifx\csname url\endcsname\relax
  \def\url#1{{\tt #1}}\fi
\expandafter\ifx\csname urlprefix\endcsname\relax\def\urlprefix{URL }\fi
\providecommand{\eprint}[2][]{\url{#2}}

\bibitem{Blote1969}
Bl\"{o}te H~W~J, Wielinga R~F and Huiskamp W~J 1969 {\em Physica\/} {\bf 43}
  549--568
  \urlprefix\url{http://www.sciencedirect.com/science/article/pii/0031891469901876}

\bibitem{Bramwell2001}
Bramwell S~T and Gingras M~J~P 2001 {\em Science\/} {\bf 294} 1495--1501
  \urlprefix\url{http://www.sciencemag.org/content/294/5546/1495.abstract}

\bibitem{Greedan2001}
Greedan J~E 2001 {\em Journal of Materials Chemistry\/} {\bf 11}(1) 37--53
  \urlprefix\url{http://dx.doi.org/10.1039/B003682J}

\bibitem{Gardner2010}
Gardner J~S, Gingras M~J~P and Greedan J~E 2010 {\em Reviews of Modern
  Physics\/} {\bf 82}(1) 53--107
  \urlprefix\url{http://link.aps.org/doi/10.1103/RevModPhys.82.53}

\bibitem{Malkin2010}
Malkin B~Z, Lummen T~T~A, van Loosdrecht P~H~M, Dhalenne G and Zakirov A~R 2010
  {\em Journal of Physics: Condensed Matter\/} {\bf 22} 276003
  \urlprefix\url{http://stacks.iop.org/0953-8984/22/i=27/a=276003}

\bibitem{Petrenko2011}
Petrenko O~A, Lees M~R and Balakrishnan G 2011 {\em Journal of Physics:
  Condensed Matter\/} {\bf 23} 164218
  \urlprefix\url{http://stacks.iop.org/0953-8984/23/i=16/a=164218}

\bibitem{Gingras2014}
Gingras M~J~P and McClarty P~A 2014 {\em Reports on Progress in Physics\/} {\bf
  77} 056501 \urlprefix\url{http://stacks.iop.org/0034-4885/77/i=5/a=056501}

\bibitem{Subramanian1983}
Subramanian M~A, Aravamudan G and Rao G~V~S 1983 {\em Progress in Solid State
  Chemistry\/} {\bf 15} 55--143
  \urlprefix\url{http://www.sciencedirect.com/science/article/pii/0079678683900018}

\bibitem{Harris1997}
Harris M~J, Bramwell S~T, McMorrow D~F, Zeiske T and Godfrey K~W 1997 {\em
  Physical Review Letters\/} {\bf 79}(13) 2554--2557
  \urlprefix\url{http://link.aps.org/doi/10.1103/PhysRevLett.79.2554}

\bibitem{Bramwell2001b}
Bramwell S~T, Harris M~J, den Hertog B~C, Gingras M~J~P, Gardner J~S, McMorrow
  D~F, Wildes A~R, Cornelius A~L, Champion J~D~M, Melko R~G and Fennell T 2001
  {\em Physical Review Letters\/} {\bf 87}(4) 047205
  \urlprefix\url{http://link.aps.org/doi/10.1103/PhysRevLett.87.047205}

\bibitem{Henley:2010vo}
Henley C~L 2010 {\em Annual Review of Condensed Matter Physics\/} {\bf 1}
  179--210

\bibitem{Fennell2009}
Fennell T, Deen P~P, Wildes A~R, Schmalzl K, Prabhakaran D, Boothroyd A~T,
  Aldus R~J, McMorrow D~F and Bramwell S~T 2009 {\em Science\/} {\bf 326}
  415--417

\bibitem{Castelnovo:2008hb}
Castelnovo C, Moessner R and Sondhi S~L 2008 {\em Nature\/} {\bf 451} 42--45

\bibitem{Castelnovo:2012kk}
Castelnovo C, Moessner R and Sondhi S~L 2012 {\em Annual Review of Condensed
  Matter Physics\/} {\bf 3} 35--55

\bibitem{Greedan1986}
Greedan J, Sato M, Yan X and Razavi F~S 1986 {\em Solid State Communications\/}
  {\bf 59} 895--897 ISSN 0038-1098
  \urlprefix\url{http://www.sciencedirect.com/science/article/pii/0038109886906526}

\bibitem{Gaulin1992}
Gaulin B~D, Reimers J~N, Mason T~E, Greedan J~E and Tun Z 1992 {\em Physical
  Review Letters\/} {\bf 69}(22) 3244--3247
  \urlprefix\url{http://link.aps.org/doi/10.1103/PhysRevLett.69.3244}

\bibitem{Zhou2008}
Zhou H~D, Wiebe C~R, Harter A, Dalal N~S and Gardner J~S 2008 {\em Journal of
  Physics: Condensed Matter\/} {\bf 20} 325201
  \urlprefix\url{http://stacks.iop.org/0953-8984/20/i=32/a=325201}

\bibitem{Ross:2011tv}
Ross K~A, Savary L, Gaulin B~D and Balents L 2011 {\em Physical Review X\/}
  {\bf 1} 021002

\bibitem{Chang:2012el}
Chang L~J, Onoda S, Su Y, Kao Y~J, Tsuei K~D, Yasui Y, Kakurai K and Lees M~R
  2012 {\em Nature Communications\/} {\bf 3} 992

\bibitem{Applegate2012}
Applegate R, Hayre N~R, Singh R~R~P, Lin T, Day A~G~R and Gingras M~J~P 2012
  {\em Physical Review Letters\/} {\bf 109}(9) 097205
  \urlprefix\url{http://link.aps.org/doi/10.1103/PhysRevLett.109.097205}

\bibitem{Robert2015}
Robert J, Lhotel E, Remenyi G, Sahling S, Mirebeau I, Decorse C, Canals B and
  Petit S 2015 {\em Physical Review B\/} {\bf 92}(6) 064425
  \urlprefix\url{http://link.aps.org/doi/10.1103/PhysRevB.92.064425}

\bibitem{Jaubert2015}
Jaubert L~D~C, Benton O, Rau J~G, Oitmaa J, Singh R~R~P, Shannon N and Gingras
  M~J~P 2015 {\em Physical Review Letters\/} {\bf 115}(26) 267208
  \urlprefix\url{http://link.aps.org/doi/10.1103/PhysRevLett.115.267208}

\bibitem{Zhou:2008cz}
Zhou H~D, Wiebe C~R, Janik J~A, Balicas L, Yo Y~J, Qiu Y, Copley J~R~D and
  Gardner J~S 2008 {\em Physical Review Letters\/} {\bf 101} 227204

\bibitem{Matsuhira2009}
Matsuhira K, Sekine C, Paulsen C, Wakeshima M, Hinatsu Y, Kitazawa T, Kiuchi Y,
  Hiroi Z and Takagi S 2009 {\em Journal of Physics: Conference Series\/} {\bf
  145} 012031 \urlprefix\url{http://stacks.iop.org/1742-6596/145/i=1/a=012031}

\bibitem{Onoda2010}
Onoda S and Tanaka Y 2010 {\em Physical Review Letters\/} {\bf 105}(4) 047201
  \urlprefix\url{http://link.aps.org/doi/10.1103/PhysRevLett.105.047201}

\bibitem{Lee2012}
Lee S, Onoda S and Balents L 2012 {\em Physical Review B\/} {\bf 86}(10) 104412
  \urlprefix\url{http://link.aps.org/doi/10.1103/PhysRevB.86.104412}

\bibitem{Kimura2013}
Kimura K, Nakatsuji S, Wen J~J, Broholm C, Stone M~B, Nishibori E and Sawa H
  2013 {\em Nature Communications\/} {\bf 4} 1934
  \urlprefix\url{http://dx.doi.org/10.1038/ncomms2914}

\bibitem{Sibille2015}
Sibille R, Lhotel E, Pomjakushin V, Baines C, Fennell T and Kenzelmann M 2015
  {\em Physical Review Letters\/} {\bf 115}(9) 097202
  \urlprefix\url{http://link.aps.org/doi/10.1103/PhysRevLett.115.097202}

\bibitem{Curnoe2008}
Curnoe S~H 2008 {\em Physical Review B\/} {\bf 78} 094418
  \urlprefix\url{http://journals.aps.org/prb/abstract/10.1103/PhysRevB.78.094418}

\bibitem{Ross2011}
Ross K~A, Savary L, Gaulin B~D and Balents L 2011 {\em Physical Review X\/}
  {\bf 1}(2) 021002
  \urlprefix\url{http://link.aps.org/doi/10.1103/PhysRevX.1.021002}

\bibitem{Hermele2004}
{Hermele} M, {Fisher} M~P~A and {Balents} L 2004 {\em Physical Review B\/} {\bf
  69} 064404
  \urlprefix\url{http://journals.aps.org/prb/abstract/10.1103/PhysRevB.69.064404}

\bibitem{Benton2012}
{Benton} O, {Sikora} O and {Shannon} N 2012 {\em Physical Review B\/} {\bf 86}
  075154
  \urlprefix\url{http://journals.aps.org/prb/abstract/10.1103/PhysRevB.86.075154}

\bibitem{Savary2012}
{Savary} L and {Balents} L 2012 {\em Physical Review Letters\/} {\bf 108}
  037202
  \urlprefix\url{http://journals.aps.org/prl/abstract/10.1103/PhysRevLett.108.037202}

\bibitem{Balakrishnan1998}
Balakrishnan G, Petrenko O~A, Lees M~R and Paul D~M 1998 {\em Journal of
  Physics: Condensed Matter\/} {\bf 10} L723
  \urlprefix\url{http://stacks.iop.org/0953-8984/10/i=44/a=002}

\bibitem{Gardner1998}
Gardner J~S, Gaulin B~D and Paul D~M 1998 {\em Journal of Crystal Growth\/}
  {\bf 191} 740--745
  \urlprefix\url{http://www.sciencedirect.com/science/article/pii/S0022024898003820}

\bibitem{Prabhakaran2011}
Prabhakaran D and Boothroyd A~T 2011 {\em Journal of Crystal Growth\/} {\bf
  318} 1053--1056
  \urlprefix\url{http://www.sciencedirect.com/science/article/pii/S0022024810010298}

\bibitem{Taguchi2002}
Taguchi Y, Ohgushi K and Tokura Y 2002 {\em Physical Review B\/} {\bf 65}(11)
  115102 \urlprefix\url{http://link.aps.org/doi/10.1103/PhysRevB.65.115102}

\bibitem{Kezsmarki2004}
K\'ezsm\'arki I, Hanasaki N, Hashimoto D, Iguchi S, Taguchi Y, Miyasaka S and
  Tokura Y 2004 {\em Physical Review Letters\/} {\bf 93}(26) 266401
  \urlprefix\url{http://link.aps.org/doi/10.1103/PhysRevLett.93.266401}

\bibitem{CiomagaHatnean2014}
{Ciomaga Hatnean} M, Decorse C, Lees M~R, Petrenko O~A, Keeble D~S and
  Balakrishnan G 2014 {\em Materials Research Express\/} {\bf 1} 026109
  \urlprefix\url{http://stacks.iop.org/2053-1591/1/i=2/a=026109}

\bibitem{Koohpayeh2014}
Koohpayeh S~M, Wen J~J, Trump B~A, Broholm C~L and McQueen T~M 2014 {\em
  Journal of Crystal Growth\/} {\bf 402} 291--298
  \urlprefix\url{http://www.sciencedirect.com/science/article/pii/S0022024814004217}

\bibitem{CiomagaHatnean2015}
{Ciomaga Hatnean} M, Lees M~R and Balakrishnan G 2015 {\em Journal of Crystal
  Growth\/} {\bf 418} 1--6
  \urlprefix\url{http://www.sciencedirect.com/science/article/pii/S0022024815000792}

\bibitem{Hanasaki2006}
Hanasaki N, Kinuhara M, K\'ezsm\'arki I, Iguchi S, Miyasaka S, Takeshita N,
  Terakura C, Takagi H and Tokura Y 2006 {\em Physical Review Letters\/} {\bf
  96}(11) 116403
  \urlprefix\url{http://link.aps.org/doi/10.1103/PhysRevLett.96.116403}

\bibitem{Kezsmarki2006}
K\'ezsm\'arki I, Hanasaki N, Watanabe K, Iguchi S, Taguchi Y, Miyasaka S and
  Tokura Y 2006 {\em Physical Review B\/} {\bf 73}(12) 125122
  \urlprefix\url{http://link.aps.org/doi/10.1103/PhysRevB.73.125122}

\bibitem{Kimura2013_1}
Kimura K, Nakatsuji S and Nugroho A~A 2013 {\em Journal of the Korean Physical
  Society\/} {\bf 63} 719--721
  \urlprefix\url{http://dx.doi.org/10.3938/jkps.63.719}

\bibitem{CiomagaHatnean2015_1}
{Ciomaga Hatnean} M, Lees M~R, Petrenko O~A, Keeble D~S, Balakrishnan G,
  Gutmann M~J, Klekovkina V~V and Malkin B~Z 2015 {\em Physical Review B\/}
  {\bf 91}(17) 174416
  \urlprefix\url{http://link.aps.org/doi/10.1103/PhysRevB.91.174416}

\bibitem{Chun2015}
Chun J, Reuvekamp P~G, Chen D, Lin C and Kremer R~K 2015 {\em Journal of
  Materials Chemistry C\/} {\bf 3}(3) 491--494
  \urlprefix\url{http://dx.doi.org/10.1039/C4TC02416H}

\bibitem{Tokiwa2014}
Tokiwa Y, Ishikawa J~J, Nakatsuji S and Gegenwart P 2014 {\em Nature
  Materials\/} {\bf 13} 356--359
  \urlprefix\url{http://dx.doi.org/10.1038/nmat3900}

\bibitem{JSY2:JSYIE5093}
Willmott P~R, Meister D, Leake S~J, Lange M, Bergamaschi A, Böge M, Calvi M,
  Cancellieri C, Casati N, Cervellino A, Chen Q, David C, Flechsig U, Gozzo F,
  Henrich B, Jäggi-Spielmann S, Jakob B, Kalichava I, Karvinen P, Krempasky J,
  Lüdeke A, Lüscher R, Maag S, Quitmann C, Reinle-Schmitt M~L, Schmidt T,
  Schmitt B, Streun A, Vartiainen I, Vitins M, Wang X and Wullschleger R 2013
  {\em Journal of Synchrotron Radiation\/} {\bf 20} 667--682 ISSN 1600-5775
  \urlprefix\url{http://dx.doi.org/10.1107/S0909049513018475}

\bibitem{Rietveld1969}
Rietveld H~M 1969 {\em Journal of Applied Crystallography\/} {\bf 2} 65--71
  \urlprefix\url{http://scripts.iucr.org/cgi-bin/paper?a07067}

\bibitem{RodriguezCarvajal1993}
Rodr\'{i}½guez-Carvajal J 1993 {\em Physica B: Condensed Matter\/} {\bf 192}
  55--69
  \urlprefix\url{http://www.sciencedirect.com/science/article/pii/092145269390108I}

\bibitem{LeBail1988}
{Le Bail} A, Duroy H and Fourquet J~L 1988 {\em Materials Research Bulletin\/}
  {\bf 23}(3) 447--452
  \urlprefix\url{http://www.sciencedirect.com/science/article/pii/0025540888900190}

\bibitem{McCusker1999}
{McCusker} L~B, {Von Dreele} R~B, {Cox} D~E, {Lou\"{e}rd} D and {Scardie} P
  1999 {\em Journal of Applied Crystallography\/} {\bf 32} 36--50
  \urlprefix\url{http://journals.iucr.org/j/issues/1999/01/00/gl0561/index.html}

\bibitem{Aharoni1998}
Aharoni A 1998 {\em Journal of Applied Physics\/} {\bf 83} 3432--3434
  \urlprefix\url{http://scitation.aip.org/content/aip/journal/jap/83/6/10.1063/1.367113}

\bibitem{Karthik2012}
Karthik C, Anderson T~J, Gout D and Ubic R 2012 {\em Journal of Solid State
  Chemistry\/} {\bf 194} 168--172
  \urlprefix\url{http://www.sciencedirect.com/science/article/pii/S0022459612003210}

\bibitem{Blanchard2013}
Blanchard P~E~R, Liu S, Kennedy B~J, Ling C~D, Avdeev M, Aitken J~B, Cowie
  B~C~C and Tadich A 2013 {\em The Journal of Physical Chemistry C\/} {\bf 117}
  2266--2273

\bibitem{CiomagaHatnean2016}
{Ciomaga Hatnean} M, Decorse C, Lees M~R, Petrenko O~A and Balakrishnan G 2016
  {\em Crystals\/} {\bf 6} 79
  \urlprefix\url{http://www.mdpi.com/2073-4352/6/7/79}

\bibitem{Sibille2016}
Sibille R, Lhotel E, {Ciomaga Hatnean} M, Balakrishnan G, F\r{a}k B, Fennell T
  and Kenzelmann M 2016 {\em Physical Review B\/} {\bf 94} 024436
  \urlprefix\url{http://journals.aps.org/prb/abstract/10.1103/PhysRevB.94.024436}

\bibitem{Anand2016}
{Anand} V~K, {Opherden} L, {Xu} J, {Adroja} D~T, {Islam} A~T~M~N,
  {Herrmannsd\"{o}rfer} T, {Hornung} J, {Sch\"{o}nemann} R, {Uhlarz} M,
  {Walker} H~C, {Casati} N and {Lake} B 2016 {\em Physical Review B\/} {\bf 94}
  144415
  \urlprefix\url{https://journals.aps.org/prb/abstract/10.1103/PhysRevB.94.144415}

\bibitem{Petit2016}
{Petit} S, {Lhotel} E, {Florea} S~G~O, {Robert} J, {Bonville} P, {Mirebeau} I,
  {Ollivier} J, {Mutka} H, {Ressouche} E, {Decorse} C, {Ciomaga Hatnean} M and
  {Balakrishnan} G 2016 {\em Physical Review B\/} {\bf 94} 165153
  \urlprefix\url{http://journals.aps.org/prb/abstract/10.1103/PhysRevB.94.165153}

\bibitem{Fukazawa2002}
Fukazawa H, Melko R~G, Higashinaka R, Maeno Y and Gingras M~J~P 2002 {\em
  Physical Review B\/} {\bf 65}(5) 054410
  \urlprefix\url{http://link.aps.org/doi/10.1103/PhysRevB.65.054410}

\bibitem{Petrenko2003}
Petrenko O~A, Lees M~R and Balakrishnan G 2003 {\em Physical Review B\/} {\bf
  68}(1) 012406
  \urlprefix\url{http://link.aps.org/doi/10.1103/PhysRevB.68.012406}

\end{thebibliography}

\end{document}